\documentclass[]{spie}  

\usepackage{amsmath,amsfonts,amssymb}
\usepackage{graphicx}
\usepackage{subfigure}
\usepackage[colorlinks=true, allcolors=blue]{hyperref}
\usepackage{svg}
\usepackage{gensymb}
\usepackage{amssymb}
\usepackage{wasysym}
\usepackage{textcomp}
\usepackage{tcolorbox}
 \usepackage{multirow}
\title{Validating the SCALES DRP: A Performance Study of the Medium-Resolution IFS Mode}

\author[a]{Athira Unni}
\author[a]{Steph Sallum}
\author[c]{Max Brodheim}
\author[c]{Rosalie McGurk}
\author[a]{Andy Skemer}
\author[c]{Matthew Brown}
\author[b]{Peyton Benac}
\author[b]{Michael P. Fitzgerald}
\author[a]{Deno Stelter}
\author[a]{Dusty Reichwein}
\author[a]{Aaron Hunter}
\author[a]{William T. S. Deich}
\author[a]{Isabel Kain}
\author[a]{Mackenzie Lach}

\affil[a]{Department of Astronomy and Astrophysics, University of California, Santa Cruz, USA}
\affil[b]{Division of Physical Sciences, University of California, Los Angeles, LA, USA}
\affil[c]{W. M. Keck Observatory, Hawai’i, USA}
\authorinfo{Corresponding Author: Athira Unni \\ E-mail: athira.exo@gmail.com }

\begin{document} 
\maketitle

\begin{abstract}
We present a performance verification of the medium-resolution integral field spectrograph (IFS) mode of the SCALES Data Reduction Pipeline (SCALES-DRP) using data obtained during recent cryogenic cooldown testing. SCALES (Slicer Combined with Array of Lenslets for Exoplanet Spectroscopy), scheduled for deployment on the Keck II Telescope in late 2026, will be the first facility-class high-contrast thermal-infrared IFS, operating over the 2--5~$\mu$m wavelength range. The SCALES-DRP, developed within the Keck Data Reduction Pipeline Framework and implemented in \texttt{Python}, converts raw detector reads into calibrated three-dimensional datacubes using rectification matrices derived from position-dependent monochromatic lenslet point-spread functions, and supports both optimal and $\chi^2$-based spectral extraction to maximize signal-to-noise and extraction fidelity. We present laboratory verification results for the medium-resolution IFS mode, which combines a $0.34^{\prime\prime}\times0.36^{\prime\prime}$ lenslet subarray with an image slicer to achieve spectral resolutions of $R\sim2,000$--$5,000$, focusing on the performance of the rectification matrices and the robustness of the extraction algorithms. These results demonstrate the readiness of the SCALES-DRP for instrument commissioning and early science operations. The pipeline is publicly available on \href{https://github.com/scalessim/SCALES-DRP}{{\ttfamily GitHub}}.
\end{abstract}


\section{INTRODUCTION} \label{sec:intro}  
SCALES (Slicer Combined with Array of Lenslets for Exoplanet Spectroscopy) is a thermal-infrared imager and integral field spectrograph (IFS) being developed for the Keck II Telescope \cite{Skemer_2022}. The instrument will operate behind the High-order Advanced Keck Adaptive Optics (HAKA) system, which provides the wavefront correction required for high-contrast observations in the thermal infrared \cite{Scott_haka}. Covering wavelengths from 2–5$\mu$m, the SCALES IFS is designed to enable direct detection and atmospheric characterization of exoplanets at temperatures down to approximately 300 K, extending ground-based high-contrast spectroscopy into a regime that is largely inaccessible to current facilities. Its high spatial resolution and contrast performance will also provide access to planetary systems at angular separations that are challenging to observe with the James Webb Space Telescope.

In addition to its exoplanet science capabilities, SCALES will support a broad range of investigations, including studies of protoplanetary disks, Solar System bodies, stellar populations, and Galactic and extragalactic sources. By combining diffraction-limited imaging with integral field spectroscopy in the thermal infrared, SCALES will open a unique region of observational parameter space and provide new opportunities for exploring planetary systems and a wide variety of astrophysical phenomena \cite{sallum_2023}.

The instrument will offer three distinct observing modes: imaging, low-spectral-resolution integral field spectroscopy, and medium-spectral-resolution integral field spectroscopy (Table \ref{tab:scales_summary}). The low-spectral-resolution IFS modes utilize a $110 \times 110$ silicon lenslet array to sample the field of view at spectral resolutions of $R \approx 35$--200. The light is dispersed by a set of six prisms. The medium-spectral-resolution mode (R $\approx$ 2000 - 5000) uses an image slicer to reformat a $17 \times 18$ lenslet subarray into a three long pseudo-slit with 102 {\ttfamily slenslit} spots each, which are then dispersed by diffraction gratings. The lengths of individual microspectra are approximately 54 pixels in the low-resolution mode and approximately 1900 pixels in the medium-resolution mode, with a separation of approximately 6 pixels between the adjacent microspectra in both modes.  Additionally, a dedicated imaging channel offers a $12.3^{\prime\prime} \times 12.3^{\prime\prime}$ field of view with a plate scale of $0.006 ^{\prime\prime}$ per pixel and a suite of 16 selectable filters spanning the $1–5~\mu$m wavelength range \cite{Banyal_2022}. SCALES includes an external calibration unit equipped with a monochromator capable of producing selectable narrowband illumination at any central wavelength between 1 and 5$\mu$m with the spectral resolution required for spectral extraction and wavelength calibration \cite{Lach_2025}. 


\begin{table}[ht]
\centering
\small
\setlength{\tabcolsep}{4pt}
\renewcommand{\arraystretch}{1.2}

\begin{tabular}{|l|ccc|ccc|c|}
\hline
 & \multicolumn{3}{c|}{\textbf{Low-Resolution IFS}} 
 & \multicolumn{3}{c|}{\textbf{Medium-Resolution IFS}} 
 & \textbf{Imager} \\
\hline

 & \textbf{Band} & \textbf{W ($\mu$m)} & \textbf{R}
 & \textbf{Band} & \textbf{W ($\mu$m)} & \textbf{R}
 &  \\
\hline

\multirow{6}{*}{\textbf{Wavelength}} 
 & K   & 2.0--2.4  & 150   & K & 2.0--2.4  & 5000 &
 \multirow{6}{*}{\begin{tabular}{c}
 Up to 16 filters\\
 1--5\,$\mu$m
 \end{tabular}} \\

 & KL  & 2.0--4.0  & 50    &   &            &      & \\

 & SED & 2.0--5.0  & 35    & L & 2.9--4.15 & 2500 & \\

 & L   & 2.9--4.15 & 80    &   &            &      & \\

 & PAH & 3.1--3.5  & 200   & M & 4.5--5.2  & 5000 & \\

 & M   & 4.5--5.2  & 200   &   &            &      & \\
\hline

\textbf{Field of View} 
 & \multicolumn{3}{c|}{2.15$\times$2.15$''$}
 & \multicolumn{3}{c|}{0.36$\times$0.34$''$}
 & 12.3$\times$12.3$''$ \\
\hline

\textbf{Spatial Sampling} 
 & \multicolumn{3}{c|}{0.02$''$}
 & \multicolumn{3}{c|}{0.02$''$}
 & 0.006$''$ \\
\hline

\end{tabular}
\caption{Summary of SCALES low-resolution IFS, medium-resolution IFS, and imager specifications \cite{Skemer_2022}.}
\label{tab:scales_summary}
\end{table}

\subsection{SCALES Data Reduction Pipeline}
SCALES employs two $2048 \times 2048$ Teledyne H2RG HgCdTe detectors, sensitive over the wavelength range of $0.6$–$5.2\mu$m. Each detector consists of a central $2040 \times 2040$ array of photosensitive pixels surrounded by a four-pixel-wide border of reference pixels \cite{Benac_2025}. The input to the SCALES Data Reduction Pipeline (SCALES-DRP) is a cube of individual detector reads with dimensions $(N \times H \times W)$, from which the pipeline produces a detector slope image and wavelength-calibrated IFS datacube.

The SCALES-DRP is implemented in {\ttfamily Python} within the Keck-DRP framework and is openly available on \href{https://github.com/scalessim/SCALES-DRP}{{\ttfamily GitHub}}, together with its documentation. The pipeline has been validated and continuously refined through multiple cryogenic laboratory test campaigns. The SCALES-DRP consists of two
primary components: (1) a calibration module and (2) a science-grade reduction module. Both the modules begin by applying a common sequence of detector-level corrections to the individual reads, including the removal of read-to-read bias variations, alternating column noise (ACN), $1/f$ noise, and amplifier-dependent channel offsets. These corrections are followed by pixel-level linearity correction and ramp fitting with jump detection to produce detector slope images. A bad-pixel correction is subsequently applied, while the linearity and bad-pixel correction together generate and propagate data-quality flags throughout the reduction \cite{unni_2025}.

The calibration module automatically processes calibration observations to generate the reference products required for science-grade data reduction. These include detector bias, dark, and flat-field exposures, together with monochromator observations for the IFS. The monochromator data are used to generate rectification matrices (RMs) for wavelength calibration and spectral extraction in each IFS observing mode, encoding the spatial and spectral response of every lenslet point spread function (PSF, or ``psflet'') across the detector. They are also used to construct lenslet flat fields and line-spread functions (LSFs). In addition, the calibration module derives a bad-pixel mask (BPM) from flat-field and dark exposures, detector gain estimation, saturation map, and linearity coefficients from flat-field sequences, read-noise maps from bias frames, and high signal-to-noise (S/N) master dark and flat files.

\begin{figure*}[!ht]
\centering
\includegraphics[width=1.0\textwidth]{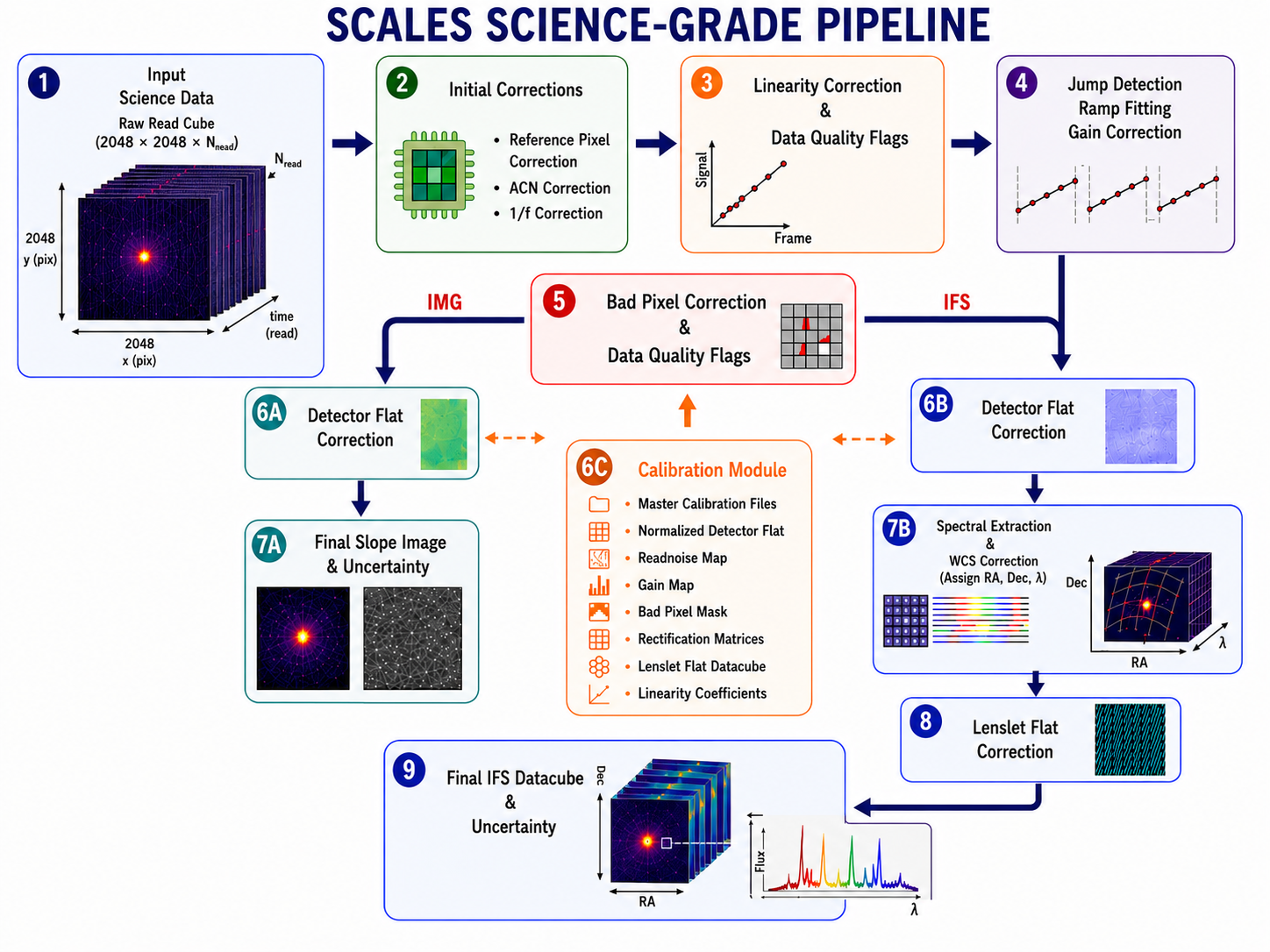}
\caption{Flowchart illustrating the science-grade reduction pipeline for both the IFS and imaging modes. The calibration module generates the calibration products required by the science-grade processing and follows the same reduction steps applied to science data \cite{unni_2025}.}
\label{flowchart_science}
\end{figure*}

The science module applies detector-level corrections, linearity correction, ramp fitting, bad-pixel correction, and spectral extraction using the calibration products generated by the calibration module. For IFS observations, the ramp-fitted slope images are transformed into wavelength-calibrated three-dimensional data cubes with associated uncertainty estimates using both optimal extraction \cite{Horne_1986} and $\chi^{2}$ extraction \cite{Brandt_2017}. Figure~\ref{flowchart_science} illustrates the overall workflow of the SCALES-DRP science module. 

An independent quicklook data reduction module operates on the SCALES server, generating detector slope images and wavelength-calibrated IFS datacubes using a subset of the science-grade reduction steps described in this paper. By processing each exposure in a fraction of a second, it provides near real-time data products to support on-sky observations. For both the imaging and IFS modes, the primary science products include detector slope images, associated uncertainty maps, and data-quality flags. For IFS observations, the pipeline additionally reconstructs a wavelength-calibrated three-dimensional data cube consisting of two spatial dimensions corresponding to the lenslet array and one spectral dimension corresponding to wavelength. Each spatial element (spaxel) contains a fully sampled spectrum.

\section{Spectral Extraction}
IFS spectral extraction transforms detector measurements into a wavelength calibrated three-dimensional datacube. The specific extraction strategy depends strongly on the instrument architecture and the distribution of spectra on the detector. One of the earliest implementations of a lenslet-based integral field spectrograph data reduction pipeline was developed for Keck/OSIRIS, introducing a rectification-matrix-based approach to spectral extraction \cite{Krabbe_2004, Lockhart_2019}. Monochromatic calibration exposures acquired using a single-column lenslet mask are used to construct empirically derived rectification matrices that characterize the detector response for each lenslet. During spectral extraction, the spatial rectification module uses these rectification matrices together with an iterative Gauss-Seidel algorithm to deconvolve the overlapping spectra and assign flux from the detector image to the corresponding spaxels \cite{Krabbe_2004, Krabbe_2006, Lockhart_2019,Samland_2022}.

Subaru/CHARIS similarly employs model-based extraction techniques, including aperture extraction, optimal extraction, and $\chi^2$-based inversion. In its $\chi^2$ extraction modes, the detector image is modeled as the forward projection of an incident spectral cube convolved with position-dependent lenslet point spread functions (PSFs), allowing least-squares optimization of lenslet fluxes while mitigating contamination from neighboring spectra \cite{Brandt_2017}. The Gemini Planet Imager (GPI) adopts an empirically calibrated microspectrum model in which wavelength-dependent microspectrum locations and PSFs are derived from calibration exposures. Positional corrections are applied to compensate for flexure and alignment shifts, after which the datacube is reconstructed using a $\chi^2$-based least-squares inversion of the detector image \cite{Draper_2014, Perrin_2014}. Likewise, the SPHERE IFS employs the BIGRE dual-microlens architecture, which reshapes the lenslet PSF to suppress diffraction wings and reduce optical crosstalk between neighboring microspectra before extraction. The original SPHERE pipeline reconstructs the datacube using calibrated microspectrum positions and aperture-based extraction, while more recent implementations additionally support optimal and least-squares ($\chi^2$) extraction using empirically measured PSFlets \cite{Antichi_2009, Beuzit_2019}.

A different detector geometry is adopted by image-slicer-based instruments. Examples include the Multi Unit Spectroscopic Explorer (MUSE) on the Very Large Telescope (VLT), the NIRSpec Integral Field Unit (IFU), and the Mid-Infrared Instrument Medium Resolution Spectrometer (MIRI MRS) aboard the James Webb Space Telescope (JWST). In these systems, the observed field is divided into narrow spatial slices that are reformatted into one or more pseudo-slits prior to dispersion, producing separated slit spectra on the detector rather than densely packed microspectra. Consequently, spectral extraction is performed on well-separated slit spectra, avoiding the significant microspectral overlap encountered in many lenslet-based IFSs. As with lenslet-based instruments, detector samples are associated with their corresponding spatial and spectral coordinates using geometric and wavelength calibrations together with an instrument optical model. The calibrated detector samples are then assembled into a three-dimensional datacube using cube-building algorithms that account for the instrument geometry and sampling pattern \cite{Weilbacher_2020, Dorner_2016, Jakobsen_2022, Wells_2015, Labiano_2021}. Reconstruction accuracy remains sensitive to slice registration, geometric distortions, wavelength calibration, and the interpolation and covariance introduced during cube building.

Collectively, these instruments illustrate the evolution of IFS spectral extraction from simple aperture summation toward increasingly sophisticated model-driven approaches. Aperture-based methods offer computational simplicity but become limited in crowded detector geometries where neighboring spectra can contaminate one another. Optimal extraction improves signal-to-noise through PSF-weighted flux estimation, while $\chi^2$ and least-squares inversion methods more completely account for detector response, spectral overlap, and field-dependent distortions. \\ In the lenslet-based architecture, the lenslet array samples the focal plane before the downstream spectrograph introduces optical aberrations, preserving the image quality delivered by the telescope and adaptive optics system. In contrast, in a slicer-based architecture, both the image slicer and the downstream spectrograph contribute optical aberrations, which can significantly degrade performance in high-contrast imaging applications where image quality is critical. The lenslet+slicer based architecture ({\ttfamily slenslit}) combines the strengths of both approaches. Like a lenslet-based IFS, it samples the focal plane before the downstream spectrograph, minimizing the impact of spectrograph aberrations on the reconstructed image. At the same time, like an image slicer, it efficiently utilizes the full detector area for spectral dispersion, enabling higher spectral resolution and detector efficiency. 

SCALES incorporates concepts from both lenslet- and slicer-based IFS architectures. The low-resolution mode operates as a conventional lenslet-based spectrograph, whereas the medium-resolution mode employs a novel lenslet-slicer architecture, referred to as a \textit{slenslit}. In this configuration, a selected lenslet subarray ($17 \times 18$) is reformatted by an image slicer into three pseudo-slits, each of which has $6 \times 17$ spots that are subsequently dispersed by diffraction gratings. The resulting spectra extend over approximately 1900 detector pixels, while mode-specific bandpass filters suppress out-of-band light and ensure that adjacent spectra remain spatially separated on the detector, minimizing direct spectral overlap.

Building upon the model-based extraction paradigm established by previous lenslet IFS instruments, SCALES utilizes empirically derived rectification matrices constructed from monochromatic calibration measurements. These matrices encode the wavelength-dependent detector response of individual spatial elements and provide the foundation for both weighted optimal extraction and global $\chi^2$ spectral extraction. In the low-resolution mode, the rectification matrix describes the detector response of individual lenslet microspectra. In the medium-resolution mode, it characterizes the detector response of the reformatted lenslet--slicer geometry. This framework enables rapid quicklook reconstruction while also supporting science-grade spectral extraction through forward modeling of the complete detector response. The implementation of these extraction methods within the SCALES Data Reduction Pipeline is described in the following sections.

\subsection{Rectification Matrix} \label{rectmat}

The rectification matrix (RM) is constructed empirically from a series of monochromatic calibration exposures that densely sample the wavelength coverage of each IFS observing mode. The SCALES calibration unit includes a monochromator capable of producing narrowband illumination at selectable wavelengths between 1 and 5~$\mu$m \cite{Lach_2025}. Each monochromatic calibration exposure produces a detector image containing the response of all illuminated spatial elements. In the low resolution mode, these appear as individual lenslet point spread functions (PSFs), commonly referred to as \textit{psflets}, while in the Medium Resolution mode they appear as spots associated with the reformatted lenslet-slicer geometry (\textit{slenslit}). Hereafter, these monochromatic detector responses are collectively referred to as spots. The number of sampled wavelength bins is chosen to adequately characterize the spectral response of each observing mode.

The RM generation process begins with automated spot detection in each monochromatic calibration image. Images are first smoothed using a Gaussian kernel to suppress pixel-scale noise, after which local maxima are identified using a peak-finding algorithm. For each wavelength plane, the centroid positions and intensities of all detected spots are recorded. These detections are then tracked sequentially through wavelength using nearest-neighbor matching, allowing the detector trajectory of each lenslet spectrum to be reconstructed across the full wavelength range. Duplicate or incomplete traces are consolidated and spurious detections are removed, producing a set of unique wavelength-dependent lenslet trajectories.

The reconstructed trajectories are subsequently registered onto a lenslet grid. Neighboring spot positions are used to recover the two-dimensional lenslet geometry and establish a one-to-one correspondence between detector coordinates and lenslet indices. This process yields a wavelength-dependent mapping between each lenslet and its associated detector footprint.

For each lenslet and wavelength sample, a small detector region surrounding the measured spot centroid is extracted directly from the calibration image. An optional local background, estimated from the outer pixels of the cropped region, is subtracted. Optionally, pixels below a fixed fraction of the psflet peak are discarded, and the remaining detector response is normalized such that its total enclosed flux is unity. Unlike analytic PSF modeling approaches, this procedure preserves the empirically measured detector response of each individual lenslet, naturally incorporating field-dependent PSF variations, optical distortions, detector sampling effects, and any residual asymmetries present in the optical system. Each normalized psflet therefore represents the detector response to unit incident flux from a specific lenslet at a specific wavelength.

The resulting RM acts as the fundamental linear forward model of the instrument, mapping flux from three-dimensional cube coordinates into detector space. Because each monochromatic spot occupies only a small detector footprint, the resulting operator is highly sparse, with non-zero values confined to the localized detector region illuminated by a particular lenslet-wavelength element. For an observing mode containing $N_{\lambda}$ wavelength channels and a lenslet field of dimensions $X_{\mathrm{lenslet}}\times Y_{\mathrm{lenslet}}$, the sparse RM has dimensions

$$
N_{\mathrm{pix}}
\times
(X_{\mathrm{lenslet}}
\times
Y_{\mathrm{lenslet}}
\times
N_{\lambda}),
$$

where $N_{\mathrm{pix}}=2048 \times 2048=4,194,304$ is the total number of detector pixels. Each column of the RM corresponds to a single spatial-spectral element and contains the normalized detector response describing how unit flux from that element is distributed across detector pixels. The matrix is stored in sparse format to enable efficient multiplication and inversion during spectral extraction. As a forward model, the RM transforms a physically meaningful three-dimensional datacube into its corresponding detector image through the linear operation $d_{\mathrm{sim}}=RA$, where $A$ is the flattened datacube flux vector, $R$ is the rectification matrix, and $d_{\mathrm{sim}}$ is the resulting detector image represented as a one-dimensional vector. Spectral extraction is formulated as the inverse problem of recovering the datacube that, when projected through the RM, best reproduces the observed detector image. The SCALES calibration framework additionally supports sub-pixel interpolation of the RM, allowing small positional and wavelength shifts arising from instrument flexure to be incorporated into the forward model without requiring a complete recalibration.


\subsection{Optimal Extraction} \label{optimal}
The SCALES pipeline implements a weighted optimal extraction framework based on the formalism introduced by Horne \cite{Horne_1986}, which has been widely adopted in astronomical spectroscopic pipelines \cite{Baranne_1996,Cushing_2004,Brandt_2017}. For SCALES, spectral extraction is formulated using the empirically derived rectification matrix, which encodes the detector response corresponding to each lenslet-wavelength element measured directly from monochromatic calibration exposures for each IFS mode. Each column of the rectification matrix is normalized to unit flux, allowing the solved amplitudes to correspond directly to reconstructed lenslet-wavelength fluxes.

For a given detector image, the observed pixel intensities are represented as a flattened detector vector $d_i$, while the rectification matrix $R$ provides the normalized detector response associated with each spatial-spectral element of the reconstructed datacube. The extracted flux for each cube element is estimated by:

\begin{equation}
F_k =
\frac{
\displaystyle\sum_i
R_{ki}\, d_i\, \sigma_i^{-2}
}{
\displaystyle\sum_i
R_{ki}^{2}\, \sigma_i^{-2}
},
\label{eq:optimal_extract}
\end{equation}

where $F_k$ is the extracted flux for the k-th lenslet-wavelength element, $R_{ki}$ is the detector response encoded in the rectification matrix for detector pixel $i$, and $\sigma^2_i$ is the total variance associated with the pixel $i$. The total pixel variance $\sigma_i^2$ is taken directly from the propagated uncertainty of the ramp-fitted slope image, which includes the combined effects of read noise, photon noise, and the ramp-fitting procedure. During extraction, the normalized detector responses are combined with the total pixel variances to form inverse-variance weights, such that detector pixels with stronger psflet responses and lower uncertainties contribute more strongly to the reconstructed flux. The rectification matrix therefore incorporates field-dependent PSF variations, detector sampling, and optical distortions directly from calibration measurements without requiring an analytic PSF model. The corresponding uncertainty in the extracted flux is propagated analytically as
\begin{equation}
\sigma_{F,k}
=
\left(
\sum_i \frac{R_{ki}^2}{\sigma_i^2}
\right)^{-1/2}
\label{eq:optimal_extract_uncertainty}
\end{equation}

The extracted flux vector is subsequently reshaped into a three-dimensional datacube with dimensions ($N_\lambda, X_{lenslet}, Y_{lenslet}$). Owing to its computational efficiency, this optimal extraction framework is used for rapid quicklook cube generation, enabling near real-time visualization of reconstructed IFS observations during data acquisition. The same extraction framework is also retained within the science-grade reduction pipeline. 

A limitation of this approach is that each spatial-spectral element is estimated independently using weighted projection onto the rectification matrix, rather than solving the full coupled detector inversion problem. As a result, flux contamination from overlapping neighboring microspectra may not be fully disentangled in crowded detector regions. While optimal extraction provides an efficient and stable reconstruction, full least-squares inversion generally yields improved flux recovery in cases where detector response overlap between adjacent lenslets becomes non-negligible.

\subsection{$\chi^2$-Based Spectral Extraction}
\label{chi_extraction}
For science-grade reductions, the SCALES pipeline implements a $\chi^2$-based spectral extraction framework that solves for the complete three-dimensional datacube simultaneously. Unlike the weighted optimal extraction described in Section~\ref{optimal}, which estimates each spatial-spectral element independently, the $\chi^2$ extraction treats the detector image as a coupled forward model of the entire lenslet field.

The extraction is formulated as the linear inverse problem

\begin{equation}
RA=d
\end{equation}

where $R$ is the sparse rectification matrix, $A$ is the unknown flux vector containing all lenslet-wavelength amplitudes, and $d$ is the observed detector image represented as a flattened vector. The best-fit solution is obtained by minimizing the variance-weighted chi-squared statistic

\begin{equation}
\chi^2=\sum_i \frac{\left[d_i-(RA)_i\right]^2} {\sigma_i^2}
\end{equation}
where $d_i$ is the observed detector signal at detector pixel $i$, $(RA)_i$ is the forward-modeled detector response, and $\sigma_i^2$ is the total variance associated with that pixel $i$.

The minimization is performed as a weighted least-squares problem by defining a diagonal weighting matrix
\begin{equation}
W=\mathrm{diag}\left(\frac{1}{\sigma_i}\right)
\end{equation}
such that the system becomes
\begin{equation}
R' = WR \qquad d' = Wd 
\end{equation}



Similar to instruments such as CHARIS \cite{Brandt_2017} and GPI \cite{Berdeu_2020}, where spectral extraction is typically performed using local detector models associated with individual microspectra or small groups of neighboring spectra, the SCALES pipeline supports both local extraction and a simultaneous global fit over the full detector image. Although the SCALES optical design provides a separation of approximately 6--7 detector pixels between adjacent microspectra in both low- and medium-resolution modes, substantially reducing direct spectral crosstalk, the global $\chi^2$ inversion further mitigates any remaining overlap through coupled forward modeling of the complete detector response.

Because the accuracy of this approach depends critically on the alignment between the rectification matrix and the observed detector geometry, SCALES will acquire one or more monochromatic calibration exposures whenever the instrument configuration is changed. These exposures will be used to measure small flexure-induced shifts and update the rectification model prior to extraction. \\
Uncertainties in the extracted datacube are estimated from the diagonal approximation to the inverse Hessian matrix of the weighted least-squares problem. After obtaining the best-fit flux vector, a model detector image is generated and used to estimate the photon-noise contribution to the detector variance. The Hessian diagonal is then approximated as
\begin{equation}
H_{jj}=\sum_i \frac{R_{ij}^2} {\sigma_i^2}
\end{equation}

yielding an estimated variance for each extracted flux element,
\begin{equation}
\mathrm{Var}(A_j) \approx \frac{1}{H_{jj}}
\end{equation}
The corresponding $1\sigma$ uncertainty is
\begin{equation}
\sigma_{A_j}= \sqrt{\mathrm{Var}(A_j)}
\end{equation}
While this approximation neglects off-diagonal covariance between neighboring spatial and spectral elements, it provides a computationally efficient uncertainty estimate that is well suited for science-grade reductions involving large sparse inversion problems. The principal advantage of this approach is its ability to simultaneously model overlapping detector responses and recover a globally self-consistent flux solution across the entire detector. The primary limitation is computational cost, as solving the full detector inversion is substantially more expensive than weighted optimal extraction and relies on efficient sparse matrix operations together with well-constrained solution boundaries.

\section{Performance Verification of Medium-Resolution Spectral Extraction}

In this section, we focus on the performance of the spectral extraction framework for the SCALES medium-resolution IFS mode. A similar demonstration for the low-resolution mode is presented in Unni et al. 2025\cite{unni_2025}. Medium-resolution IFS pinhole spectra and monochromatic calibration exposures for the K-, L-, and M-band modes were obtained during the most recent cryogenic cooldown conducted at the UCSC laboratory beginning on 2026 June 10. Here, we present the K-band spectral extraction results in detail.

\begin{figure*}[!ht]
\centering
\includegraphics[width=\textwidth]{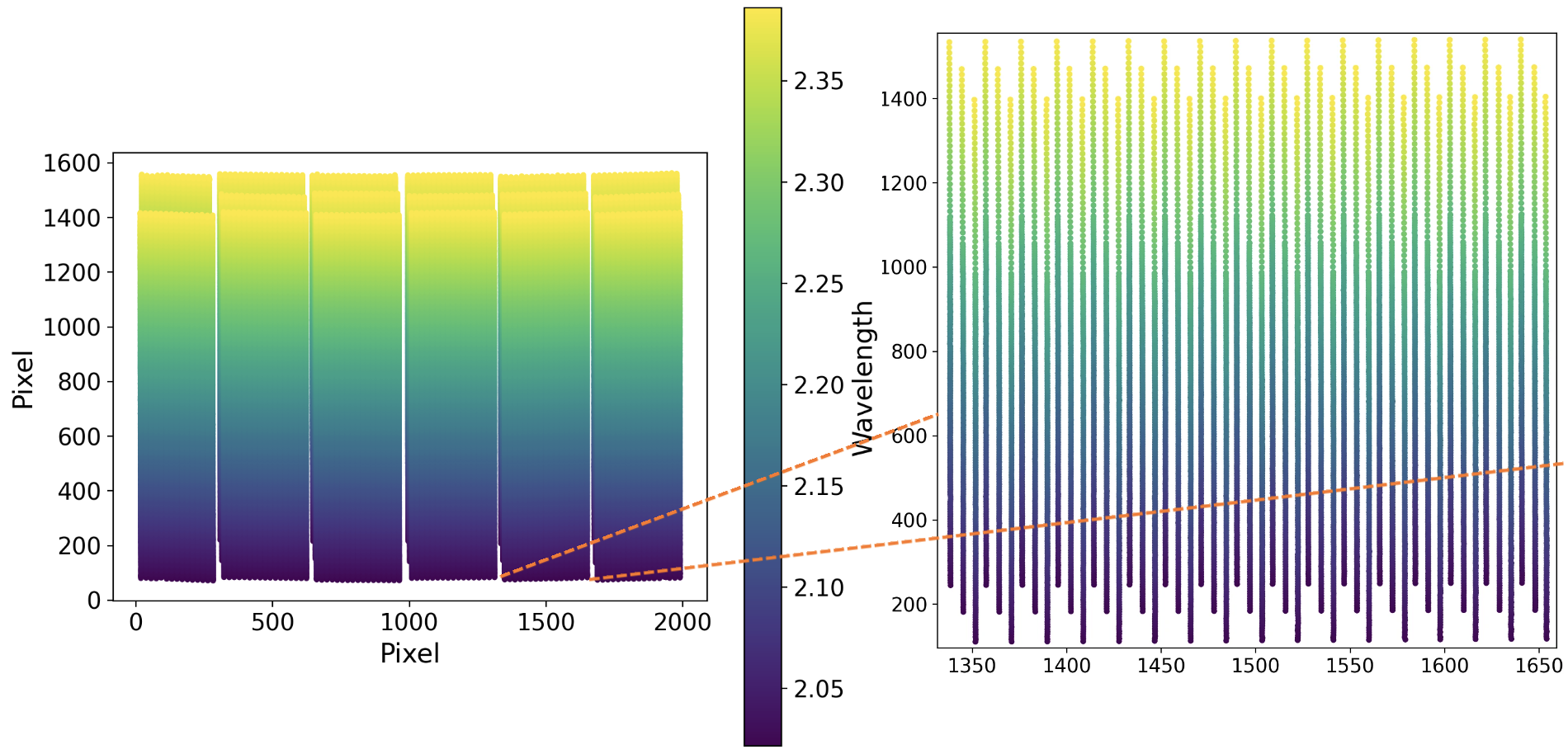}
\caption{Measured detector trajectories of the lenslet spots in the SCALES medium-resolution K-band mode, traced across the full wavelength range from 2.0 to 2.4~$\mu$m, with wavelength increasing from bottom to top. Six groups of traces are visible across the detector, each containing 51 individual spectra. The spectra extend approximately 1500 detector pixels in the dispersion direction and are separated by $\sim$7 pixels within each group, while adjacent groups are separated by $\sim$31 pixels. This detector geometry minimizes spectral overlap and forms the basis for constructing the medium-resolution rectification matrix used for spectral extraction. The spectral extent shown along the detector $y$-direction is based on the K-band filter installed in the imager filter wheel, as the dedicated IFS K-band filter was unavailable during the cooldown campaign. Consequently, the spectral extent will differ from that of the final on-sky configuration when the flight IFS K-band filter is used.}
\label{k_trace}
\end{figure*}

For the K-band calibration, monochromatic exposures spanning the wavelength range of $2.0$--$2.4,\mu$m are acquired to create the rectification matrix. The wavelength scan is sampled at an effective spectral resolution of $R \sim 500$, with each wavelength point consisting of a combination of nine 20s exposures, yielding a signal-to-noise ratio of approximately 16. All the basic data reductions including detector level corrections, linearity correction, jump detection, ramp fitting, and bad pixel correction are performed. Figure \ref{k_trace} shows the measured detector trajectories of the \textit{slenslit} spots across the full wavelength range and illustrates the distribution of the medium-resolution spectra on the detector.

\begin{figure*}[!ht]
\centering
\includegraphics[width=0.9\textwidth]{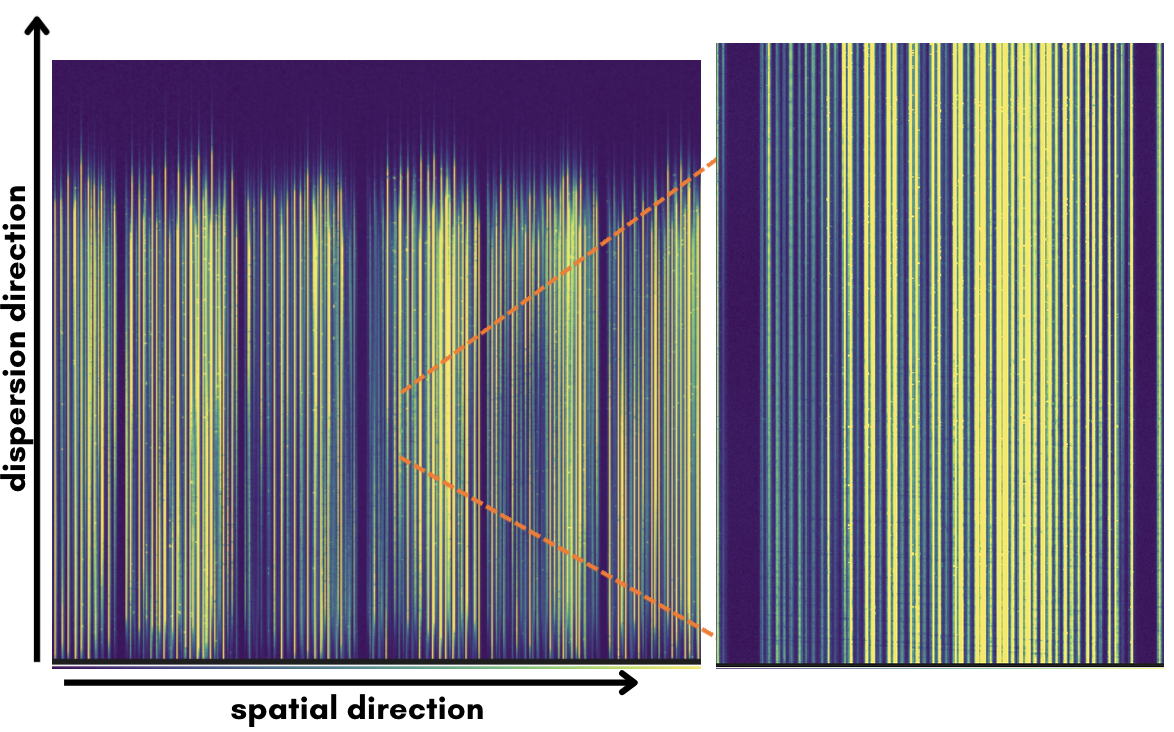}
\caption{Background-subtracted medium-resolution K-band pinhole spectra obtained with the SCALES K-band grating and filter combination. Six groups of spectra are visible across the detector, each containing 51 individual spectra generated by the {\ttfamily slenslit} optics. The spectra are approximately 1500 pixels long in the dispersion direction and are separated by $\sim$7 pixels within each group, while adjacent groups are separated by $\sim$31 pixels.}
\label{k_2d_dis}
\end{figure*}

\begin{figure*}[!ht]
\centering
\includegraphics[width=\textwidth]{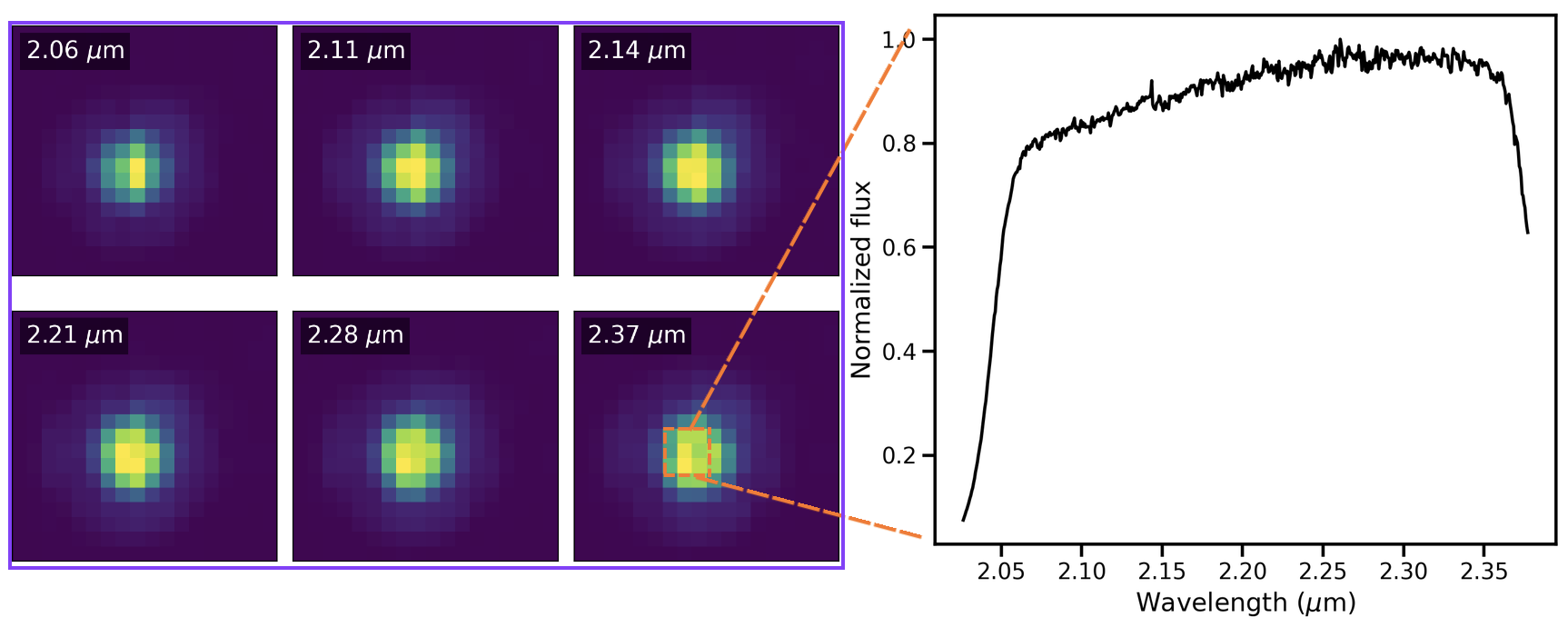}
\caption{The medium resolution K-band IFS cube slice at different wavelengths from 2.0-2.4 $\mu$m and a one dimensional extracted spectra integrated over the square region.}
\label{k_cube_spectra}
\end{figure*}

To evaluate the extraction performance, dispersed pinhole data were acquired using the combination of the K-band diffraction grating in the IFS filter wheel and the K-band filter in the imager filter wheel. Because the dedicated K-band filter for the IFS filter wheel was unavailable during the cooldown campaign, the spectral extent along the dispersion direction differ from those expected during normal science observations with the nominal IFS K-band filter. The final detector image was produced by combining nine 15s exposures and subtracting the corresponding background frames. The resulting detector image is shown in Figure~\ref{k_2d_dis}. Six groups of spectra are visible on the detector, with adjacent groups separated by approximately 31 pixels. Each group contains 51 individual spectra arranged in a non-overlapping pattern and dispersed into traces approximately 1500 detector pixels in length. Neighboring spectra within a group are separated by approximately 7 pixels, minimizing direct spectral overlap and enabling robust spectral extraction through the rectification-matrix framework described in Section \ref{rectmat}. Figure~\ref{k_cube_spectra} shows representative slices of the extracted datacube at different wavelengths, together with the corresponding one-dimensional spectrum obtained using the optimal extraction. 

\section{Conclusion}
SCALES is a thermal-infrared integral field spectrograph and diffraction-limited imager currently on track for first light on the Keck II Telescope in late 2026. The SCALES Data Reduction Pipeline (SCALES-DRP) is being developed to provide both rapid quicklook analysis and science-grade data products for imaging and IFS observations. For imaging observations, the primary science product is a calibrated detector slope image together with the associated data quality flags and uncertainty products. For IFS observations, the pipeline additionally reconstructs a three-dimensional datacube consisting of two spatial dimensions and one spectral dimension. To support a wide range of observing needs, SCALES-DRP provides both computationally efficient quicklook reductions and science-grade processing that includes advanced spectral extraction techniques.

Recent cryogenic cooldown testing at UCSC successfully verified the medium-resolution IFS reduction and spectral extraction framework, including the generation and application of rectification matrices and the reconstruction of calibrated datacubes. These results represent an important milestone toward instrument commissioning and demonstrate the readiness of the medium-resolution data reduction pipeline for on-sky operations and early science observations.

\acknowledgments

Major support for the SCALES project has been provided through NSF Grant No. 2216481, as well as grants from the Heising-Simons Foundation, the Mt. Cuba Astronomical Foundation, and the Alfred P. Sloan Foundation. We are also grateful to the Robinson family and other private supporters, whose generosity has been instrumental in making this work possible.

\normalsize
\setlength{\baselineskip}{0.8\baselineskip}
\bibliography{report} 
\bibliographystyle{spiebib} 

\end{document}